\documentclass[aps,pra,twocolumn,showpacs,longbibliography,superscriptaddress,floatfix]{revtex4-1}
\usepackage[colorlinks=true, allcolors=blue]{hyperref}
\usepackage{amsmath}
\usepackage{dsfont}
\usepackage{amssymb}
\usepackage{braket}
\usepackage{color}
\usepackage{eucal}
\usepackage{xcolor}
\usepackage{subfig}
\usepackage{caption}
\usepackage{graphicx}
\usepackage{mathdots} 
\usepackage{booktabs}
\newcommand{\la}{\langle}
\newcommand{\ra}{\rangle}

\newcommand{\quotes}[1]{``#1''}
\newcommand{\singquotes}[1]{`#1'}
\usepackage{wrapfig}    % for wrapfigure environment

\begin{document}

\title{Uncertainty Relation for Non-Hermitian Systems}
\author{Namrata Shukla}
\affiliation{Department of Physics, Institute of Science, BHU, Varanasi 221005, India}
\author{Ranjan Modak}
\affiliation{Department of Physics, Indian Institute of Technology Tirupati, Tirupati 517619, India}
\author{Bhabani Prasad Mandal}
\affiliation{Department of Physics, Institute of Science, BHU, Varanasi 221005, India}

\begin{abstract}
%We investigate the validity of the Heisenberg uncertainty relation for $\mathcal{PT}$ invariant non-Hermitian quantum theories.
%In the usual quantum theory, 
%Robertson's formalized version of the Heisenberg uncertainty relation contains a state of interest and two incompatible observables that are Hermitian operators.
%Our construction is not limited to the $\mathcal{PT}$-symmetric phase but also valid in the $\mathcal{PT}$-broken phase.}
%\nam{The fact that the non-Hermitian Hamiltonian itself is a good observable can be used to diagnose the $\mathcal{PT}$ phase transition.
We construct uncertainty relation for arbitrary finite dimensional $\mathcal{PT}$ invariant non-Hermitian quantum systems within a special inner product framework.~This construction is led by \quotes{good observables} which are a more general class of operators.~We show that the cumulative gain in the quantum Fisher information when measuring two good observables for such non-Hermitian systems is way better than their Hermitian counterpart.~Minimum uncertainty states being the best candidates for this gain near the exceptional point supports the intelligent or simultaneous non-Hermitian quantum sensors.

%For actual measurement purposes, we construct two incompatible Hermitian good observables and identify the quantum states that minimize the uncertainty in the measurement of these observables, near the exceptional point (EP).~These special quantum states we call \quotes{exceptional states} coincide with the minimum uncertainty states, near the EP.~We suggest a method to use almost simultaneous measurement for strong quantum sensing near the EP.

%demonstrate this  for the general  $2\times2$ PT-symmetric non-Hamiltonian Hamiltonian and introduce a generalized approach to the same for arbitrary finite dimension.
\end{abstract}
\maketitle
\section{Introduction}
Heisenberg uncertainty principle~\cite{Heisenberg_1927,Robertson_1929,Wheeler_1983} is one of the most important tenets of quantum mechanics.~It encapsulates the impossibility of simultaneous measurement of two incompatible observables.~In the standard quantum mechanics, the real spectrum, complete set of orthonormal eigenstates with the positive definite norm and  unitary evolution of the system are guaranteed by Hermitian Hamiltonian.~Operators for the physical observables are Hermitian with the measurement result as one of its real eigenvalues.~The Heisenberg uncertainty relation rigorously proved by Robertson~\cite{Robertson_1929} for the two general incompatible observables is  \begin{equation}
\Delta A^2\Delta B^2\geq \frac{1}{4}{\vert\bra{\psi}[A,B]\ket{\psi}\vert}^2.
\label{Heis-Rob}
\end{equation}~Uncertainty relation has been actively debated in the last two decades under the two different outlooks of \singquotes{preparation} and \singquotes{measurement} \cite{Uffink_1985, Erhart_2012, Mukhopadhyay_2016, Pati_2014, Pati_2014,Shukla_2018} followed by various experimental realizations \cite{Rouze_1993,Xiao_2017,Qu_2021}.

Inequality \eqref{Heis-Rob} saturates for the minimum uncertainty states (MUS). MUS are vital because these states are expected to reproduce, as closely as possible, the classical motion, and those 
%~An example of MUS is coherent state which was initially introduced for the Harmonic oscillator but later have been generalized for an arbitrary potential \cite{ray_1982,nieto_1978}.
 have been of utmost importance in the broad area of physics starting from quantum optics \cite{harry_1972} to the theoretical developments in quantum gravity \cite{pierre_2021}.~If one defines the ratio of LHS and RHS of the inequality \eqref{Heis-Rob} as $\eta$, then $\eta=1$ corresponds to the MUS.~It has also been shown that the MUS are the eigenstates of the operator $A+i\lambda B$ for $\lambda=\Delta A\slash\Delta B$ \cite{Barone_2021}.
%The inequality\eqref{Heis-Rob}can also be simplistically written as  
%\begin{equation*}
%\eta=\frac{2\Delta A^2\Delta %B^2}{{\vert\bra{\psi}[A,B]\ket{\psi}\vert}^2}. 
%\end{equation*}
%Violation of the uncertainty relation \eqref{Heis-Rob} implies %$\eta<1$. The minimum uncertainty states (MUS) as the name %suggests saturate the uncertainty inequality \eqref{Heis-Rob} %implying $\eta=1$. It has also been shown that these MUS states %are the eigenstates of the operator$A+i\lambda B$ for %$\lambda=\Delta A\slash\Delta B$ \cite{Barone_2021}.\\

%\nam{Do we need to revisit and shorten this paragraph?}
Over the past two decades, there has been great interest in a certain class of non-Hermitian quantum theories where 
Hermiticity condition on the Hamiltonian of the system is replaced with a physical and rather less constraining condition of $\mathcal{PT}$-symmetry~\cite{Bender_1998, Bender_2002,Bender_2007, Zhou_2020,Klauck_2019,Das_2010,Ashida_2017,Khare_2000, Zhang_2019,Bagarello_2018,Bagarello_2021}.~Such $\mathcal{PT}$-invariant non-Hermitian systems generally exhibit a $\mathcal{PT}$ symmetry breaking transition that separates two  regions, (i) $\mathcal{PT}$-symmetric phase in which the entire spectrum is real and the eigenfunctions of the Hamiltonian respect $\mathcal{PT}$ symmetry and (ii) $\mathcal{PT}$-broken phase 
%\nam{Would it not be nice to label here the %$\mathcal{PT}$-symmetric and $\mathcal{PT}$-broken phase.}
in which the entire spectrum (or a part of it) is in complex conjugate pairs and the eigenstates of the Hamiltonian are not the eigenstates of $\mathcal{PT}$ operator.~The phase transition happens at the exceptional point (EP) for the particular Hamiltonian.
~It has been well established that 
 a modified Hilbert space equipped with a positive definite $\mathcal{CPT}$ inner product \cite{Bender_2002}
 can lead to consistent quantum theories  in the
 $\mathcal{PT}$-symmetric phase, but very little has been explored in the broken phase which also plays a significant role in developing non-Hermitian theories
 \cite{Bender_2007,Zhou_2020,Klauck_2019,Das_2010,Ashida_2017, Ozdemir_2019}.~While the $\mathcal{CPT}$ inner product is restricted to the $\mathcal{PT}$-symmetric phase only, recently proposed~$G$-metric inner product based on the geometry of the quantum states by defining a connection-compatible positive definite metric operator~$G$~ %which is positive definite
\cite{Mostafazadeh_2007, Ju_2019, Gardas_2016,Huang_2021}, can be extended even in the $\mathcal{PT}$-broken phase.

\begin{figure}
\centering
\includegraphics[width=0.8\linewidth]{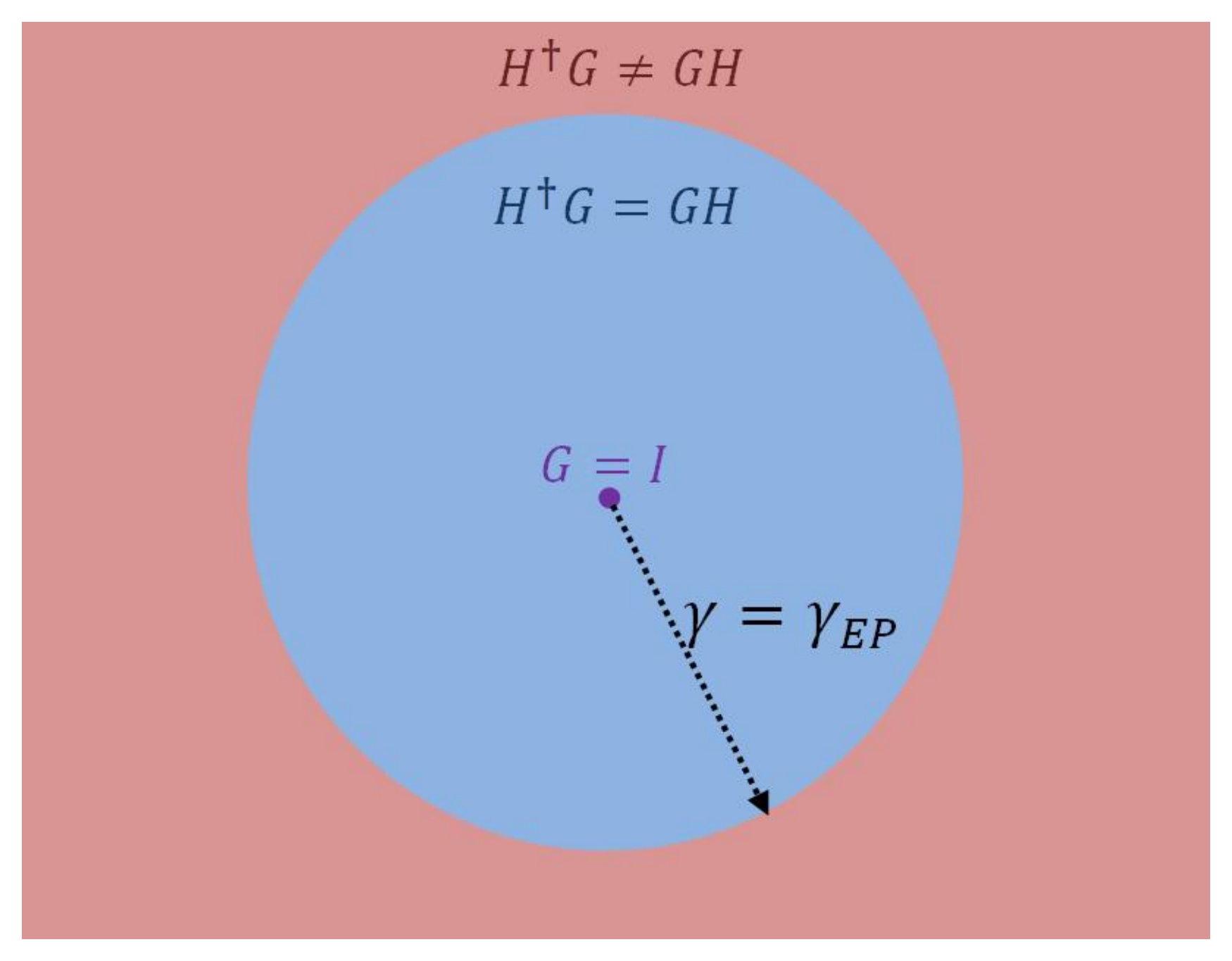}
\caption{General pictorial representation of the \quotes{good observable} condition on the Hamiltonian in the $\mathcal{PT}$-symmetric and $\mathcal{PT}$-broken region with the exceptional point $\gamma=\gamma_{EP}$.~The condition reduces to Hermiticity at the center with $G=\mathds{1}$} and $\gamma=0$.
\label{Figure0}
\end{figure}

The question arises, what is the fate of the Heisenberg uncertainty relation 
for  the system described by $\mathcal{PT}$-invariant non-Hermitian Hamiltonian?~This question is critical because the behaviour of uncertainty relation can be highly validating of the newly proposed theory of $\mathcal{PT}$-symmetric non-Hermitian quantum mechanics.~The uncertainty relation for position and momentum for such non-Hermitian systems has been discussed in a rather different context for a continuum model \cite{Bagchi2009}.
In this work, we show that  the Hermiticity condition on the operator $O$ can be replaced by a more general condition we call \quotes{good observable} meaning $O^{\dag}G=GO$ and use it to construct the modified uncertainty relation for two such good observables $A$ and $B$ 
\begin{equation}
\Delta A^2_G\Delta B^2_G\geq \frac{1}{4}{\vert\bra{\psi}G[A,B]\ket{\psi}\vert}^2,
\label{Heis-Rob_G}
\end{equation}
for an arbitrary finite dimensional system.~Here, $\Delta A_G^2=\la \psi|GA^2|\psi\ra -(\la \psi|GA|\psi\ra)^2$ and $\Delta B_G^2=\la \psi|GB^2|\psi\ra -(\la \psi|GB|\psi\ra)^2$ and $G$ is a Hermitian positive definite matrix.
~A good observable can be either Hermitian or non-Hermitian and the uncertainty relation is valid in the $\mathcal{PT}$-symmetric as well as $\mathcal{PT}$-broken phase.~We also show that the non-Hermitian Hamiltonian itself is one of such good observables in the $\mathcal{PT}$-symmetric phase but not in the $\mathcal{PT}$-broken phase, thus indicating the EP for the system.~The pictorial representation of the good observable condition on the Hamiltonian in Fig.~\ref{Figure0} clearly demonstrates that the usual Hermitian quantum mechanics is a special case of $\mathcal{PT}$-symmetric non-Hermitian quantum mechanics under the G-metric inner product formalism in the context of the observables.~Also, MUS in this context are represented by $\eta_G=1$ for $\eta_G$ defined as the ratio of the LHS and RHS of the Eq.~\ref{Heis-Rob_G} and $\eta_G= \eta $ for Dirac inner product $(G=\mathds{1})$.~Violation of the uncertainty relation implies $\eta_G<1$ for Eq.~\eqref{Heis-Rob_G} and $\eta<1$ for Eq.~\eqref{Heis-Rob}.

Quantum Fisher information (QFI) indicates the sensitivity of a state with respect to the perturbation of parameters and hence used in quantum metrology~\cite{Giovannetti2011}.~The gain in QFI implies the enhancement in the precision of parameter estimation and it has been an active area of investigation~\cite{Giovannetti2011, Wang2020, Rath_2021, Yu_2020}.~We identify Hermitian good observables for the non-Hermitian systems and define the cumulative gain in Fisher information with respect to its Hermitian equivalent when measuring energy with this Hermitian good observable.~We then show that this cumulative gain shows a drastic improvement near the EP for the MUS over the random states.~We propose to use this feature as a candidate for the most precise simultaneous measurement leading to simultaneous quantum sensing near the EP for such systems \cite{Ma2022}.

%%%%%BPM%%%%%
\section{$2\times2$ Non-Hermitian System}
To realize the modified uncertainty relation, we begin with a $2\times2$ model of one parameter $\mathcal{PT}$-invariant system described by the Hamiltonian
\begin{equation}
    H(\gamma)=
    \begin{bmatrix} 
	i\gamma & 1  \\
	1 & -i\gamma\\
	\end{bmatrix},
	\quad
	\label{Hamiltonian}
\end{equation}
whose eigenvectors are denoted as $\ket{E_1}$ and 
$\ket{E_2}$.~By tuning the parameter $\gamma$, one can go from $\mathcal{PT}$-symmetric phase to the $\mathcal{PT}$-broken phase, the EP is at $\gamma=\gamma_{EP}=1$.

%%%%%%%%%%%%%%%%%%%%%%%%%%%%%%%%%%%%%%%%%%%%%%%%%%%%%%%%%%%%%%%%%%%%%%%%%%%%
\begin{figure}
\centering
\includegraphics[width=1\linewidth]{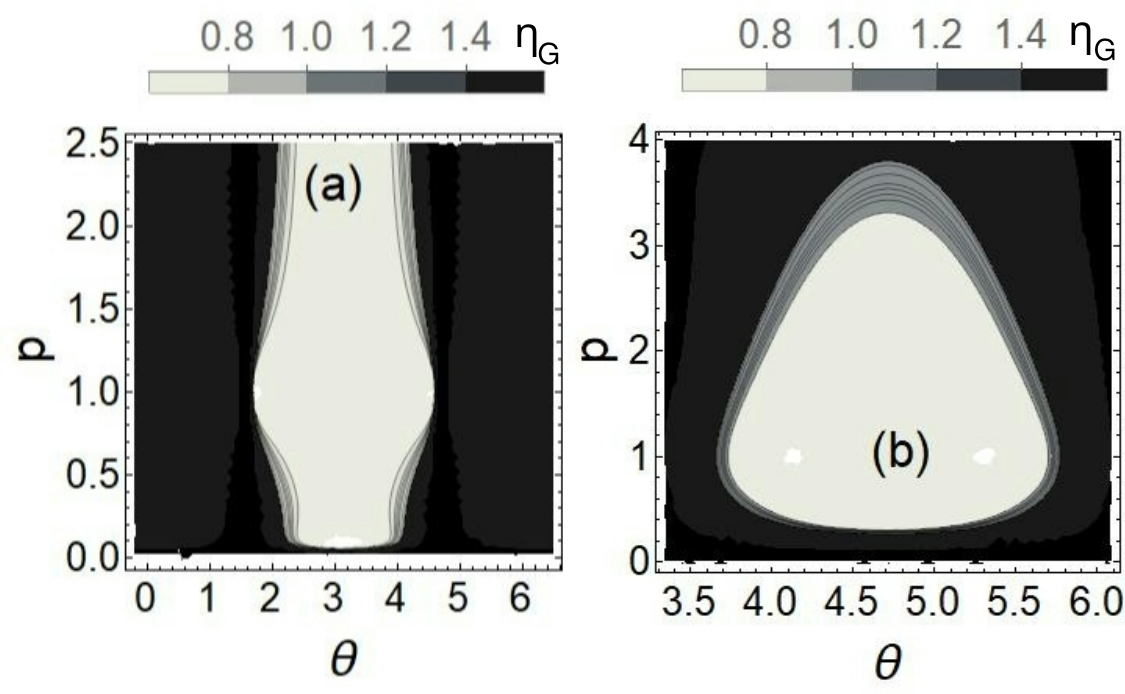}
\caption{Contour plots for $\eta_G$ for observables $\sigma_x$ and $\sigma_z$ and generic state \eqref{genstate} in (a) $\mathcal{PT}$-symmetric phase ($\gamma=0.2$), and (b) $\mathcal{PT}$-broken phase ($\gamma=1.2$).~Violation is seen because they are not good observables in the $G$-metric inner product framework.}
\label{Figure1}
\end{figure}
%%%%%%%%%%%%%%%%%%%%%%%%%%%%%%%%%%%%%%%%%%%%%%%%%%%%%%%%%%%%%%%%%%%%%%%%%%%%%%
\begin{figure}
\centering
\includegraphics[width=1\linewidth]{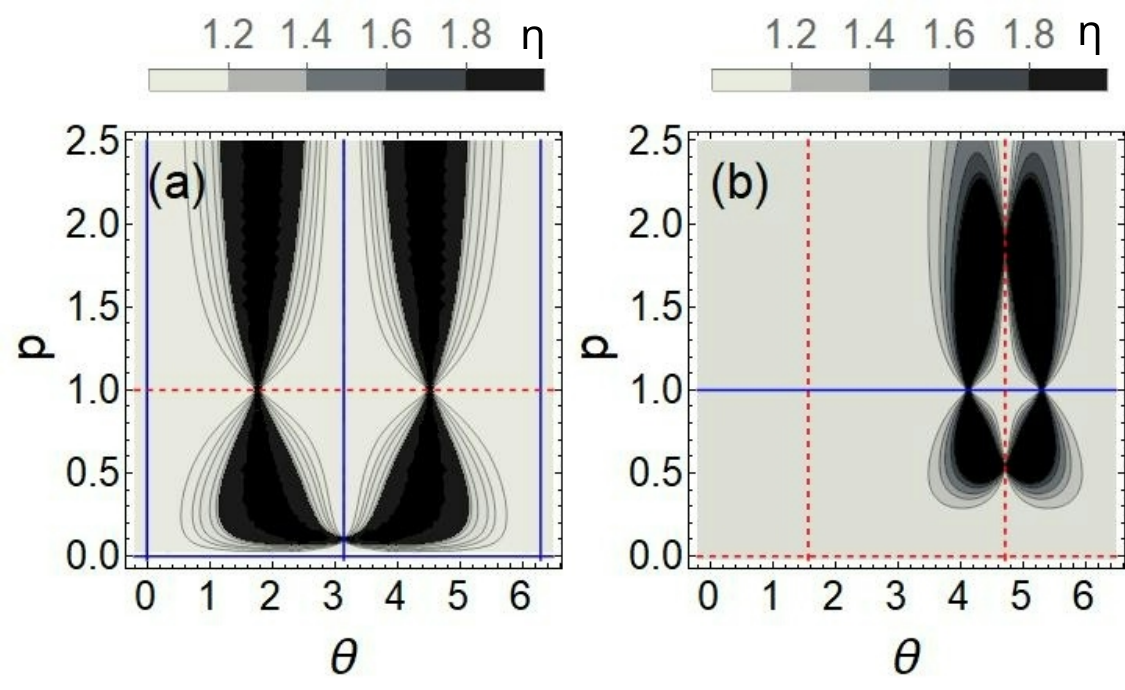}
\caption{Contour plots for $\eta$ for the observables $\sigma_x$ and $\sigma_z$ and generic state \eqref{genstate} in (a) $\mathcal{PT}$- symmetric phase ($\gamma=0.2$), and (b) $\mathcal{PT}$-broken phase ($\gamma=1.2$).~Solid blue lines correspond to $\mathcal{PT}$ symmetric MUS and red dashed lines correspond to  $\mathcal{PT}$ broken MUS.}
\label{Figure2}
\end{figure}
%%%%%%%%%%%%%%%%%%%%%%%%%%%%%%%%%%%%%%%%%%%%%%%%%%%%%%%%%%%%%%%%%
\begin{figure}
\centering
\includegraphics[width=1\linewidth]{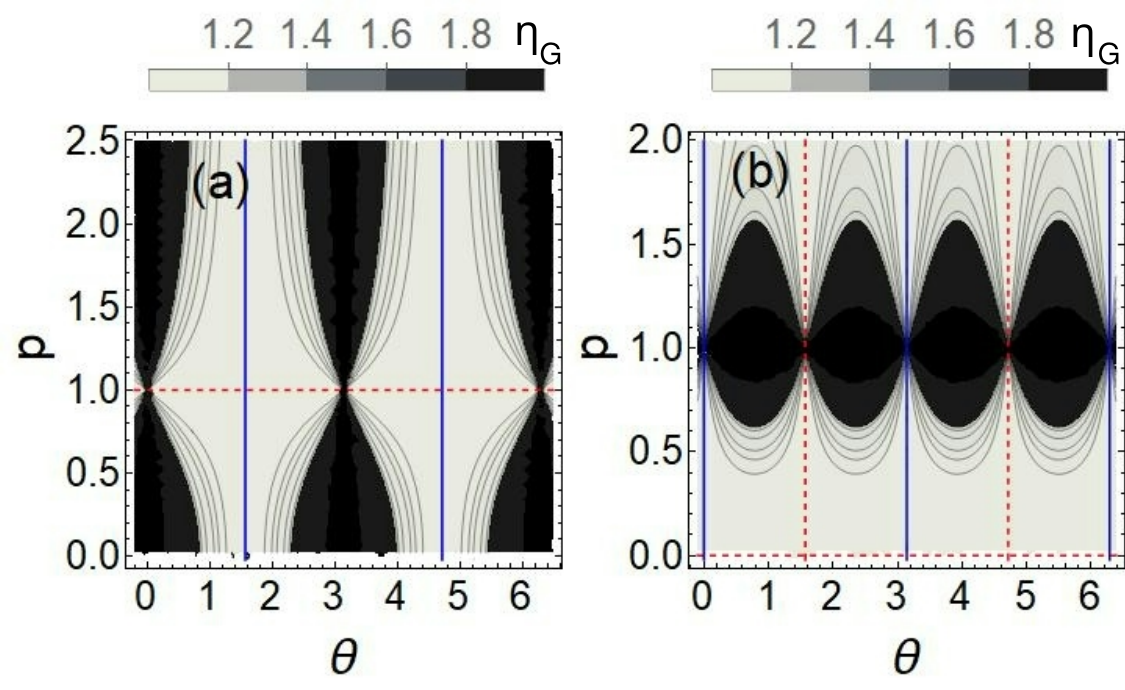}
\caption{Contour plots for $\eta_G$ and the generic state \eqref{genstate} in (a) $\mathcal{PT}$-symmetric phase ($\gamma=0.2$) for the good observables $H(\gamma)$ and $\sigma_y$, and (b) $\mathcal{PT}$-broken phase ($\gamma=1.2$) for the good observables $H(1/\gamma)$ and $\sigma_y$.~Solid blue lines correspond to $\mathcal{PT}$ symmetric MUS and red dashed lines correspond to $\mathcal{PT}$ broken MUS.} 
\label{Figure3}
\end{figure}

A state is called $\mathcal{PT}$ symmetric state if it respects the $\mathcal{PT}$ symmetry and $\mathcal{PT}$ broken state otherwise.~Thus in the $\mathcal{PT}$-symmetric phase $\ket{E_{1,2}}$ are also eigenstates of $\mathcal{PT}$ operator.~The state we choose to demonstrate the behaviour of uncertainty relation is a general superposition of the eigenstates of the Hamiltonian $H$ in both the $\mathcal{PT}$-symmetric and $\mathcal{PT}$-broken phase
\begin{equation}
\ket{\Psi}=\mathcal{N}(\ket{E_1}+p\text {e}^{i\theta}\ket{E_2}),\label{genstate}
\end{equation}
where $\mathcal{N}$ is the normalization constant and $p, \theta$ are the state parameters.

For the non-Hermitian 
$\mathcal{PT}$-invariant Hamiltonian, it is necessary to define a new inner product that 
%not only reproduces the bi-orthogonal quantum mechanics but also 
makes the eigenvectors of this Hamiltonian orthonormal.~The inner product defined with a positive definite metric operator $G$ satisfies this property and can be chosen as $G=\sum_i\ket{L_i}\bra{L_i}$,~$\ket{L_i}$ being the $i^{\text th}$ left-eigenvector of $H$.~The dual vector of $\ket{\psi}_G$ in this formalism is defined as $\bra{\psi}_G=\bra{\psi}G$ and for the corresponding $\ket{\psi}_G$, the inner product reads $\la{\psi}\ket{\psi}_G=\bra{\psi}G\ket{\psi}$.~The matrix $G$ in general is time-dependent and satisfies the equation of motion
\begin{equation*}
     \partial_tG=i[GH-H^{\dag}G]. \nonumber
\end{equation*}
In the $\mathcal{PT}$-symmetric phase $\partial_tG=0$, and it implies that the $G$ is time-independent.~However, in the $\mathcal{PT}$-broken phase, $G$ is time-dependent.~The matrix form of $G$ for the $2\times2$ systems in the $\mathcal{PT}$-symmetric and $\mathcal{PT}$-broken phase at $t=0$, respectively takes the form
\begin{equation}
G^s =\frac{1}{\sqrt{1-\gamma^2}}
   \begin{bmatrix} 
	1 &- i\gamma  \\
	i\gamma & 1\\
	\end{bmatrix},
	\quad
    G^b =\frac{1}{\sqrt{\gamma^2-1}}
    \begin{bmatrix} 
	\gamma & -i \\
	i & \gamma\\
	\end{bmatrix}.
	\quad
	\label{G-symmetric_broken}
\end{equation} 
%for $t=0$ and $\lambda=\sqrt{\gamma^2-1}$.

In the usual quantum mechanics, the real expectation value of operators imposes the \singquotes{Hermiticity} condition on the operators, i.e., $O=O^{\dag}$.~However, in the non-Hermitian quantum mechanics equipped with $G$-inner product, making the $G$-expectation value of an operator $O$ for any state $\ket{\psi}$ real leads to  
\begin{equation} 
O^{\dag}G=GO.    
\label{good obs}
\end{equation}
We coin the term \quotes{good observable} for the operators that satisfy this condition.~Since $G$ is Hermitian, for Hermitian operators the condition \eqref{good obs} automatically means $[G, O]=0$.~For the Dirac inner product ($G=\mathds{1}$) the condition \eqref{good obs} trivially reduces to the Hermiticity condition on the operators. It is also straightforward to show from Eq.~\eqref{good obs} that there exists 
a similarity transformation that maps a non-Hermitian good observable $O$ to a Hermitian operator $\tilde{O}=G^{1/2}OG^{-1/2}$.~Obviously, all the Hermitian operators are good observables for the Dirac product.~However, for the non-Hermitian systems, what will be a good observable will depend upon the $G$-metric and it can have different structures in the $\mathcal{PT}$-symmetric
and $\mathcal{PT}$-broken phase.~The structure of good observables in the $\mathcal{PT}$-symmetric phase ($\gamma <1)$ is
\begin{equation}
O^{s}=
\begin{pmatrix}
i\gamma x & x-iy \\
x+iy & -i\gamma x
\end{pmatrix},
\label{go_sym}
\end{equation}
and in the $\mathcal{PT}$ broken phase ($\gamma >1)$ for $t=0$ is
\begin{equation}
O^{b}=\frac{1}{\gamma}
\begin{pmatrix}
ix & \gamma(x-iy) \\
\gamma(x+iy) & -ix
\end{pmatrix},
\label{go_broken}
\end{equation}
where $x$, $y$ are real numbers.
We like to emphasize that the structure of good observables remains the same at all times in the $\mathcal{PT}$-symmetric phase (since the $G$-metric is time-independent in this phase), but not in the $\mathcal{PT}$-broken phase.~The one mentioned in Eq.~\eqref{go_broken} is the structure of good observable in the $\mathcal{PT}$-broken phase at the initial time $t=0$.~The time dependence of $G$ and hence the good observables is discussed in the Appendix \ref{appendix F}.~To test our new uncertainty relation, one can choose the observables $\sigma_x$ and $\sigma_z$ for the generic state \eqref{genstate} with the $G$-metric inner product.~The uncertainty relation \eqref{Heis-Rob_G} shows violation for a rather big segment of states for both $\mathcal{PT}$- symmetric and $\mathcal{PT}$-broken phase in Fig.~\ref{Figure1}.~Clearly, despite being Hermitian operators $\sigma_x$ and $\sigma_z$ are not good observables for the $G$-metric inner product.~However, it is worth mentioning that within the Dirac product, one can choose Hermitian pairs of non-commuting observables $\sigma_x$ and $\sigma_z$ over the state \eqref{genstate} and see no violation of the uncertainty relation \eqref{Heis-Rob} for any value of the state parameters, as anticipated (see Fig.~\ref{Figure2}).
%~It still remains a challenge to find time-independent good observables in the broken phase which remain good observables for all time (see Appendix F for further discussion).}
%\nam{We learned at a later point that the uncertainty relation for position and momentum for such non-Hermitian systems has been discussed in a rather different context \cite{Bagchi2009}}.

For the non-Hermitian systems, an example of the set of two incompatible good observables is $H(\gamma)$, $\sigma_y$ in the $\mathcal{PT}$-symmetric and $H(1/\gamma)$, $\sigma_y$ in the $\mathcal{PT}$-broken phase.~As evident from Fig.~\ref{Figure3}, the uncertainty relation \eqref{Heis-Rob_G} holds good as long as one chooses the good observables.
%Note that $H(\gamma)$ does not satisfy the good observables condition in Eq.~\eqref{good obs} for the $\mathcal{PT}$-broken phase.

%%%%%%%%%%%%%%%%%%%%%%%%%%%%%%%%%%%%%%%%%%%%%%%%%%%%%%%%%

In usual quantum mechanics, MUS for the two incompatible observables $A$ and $B$ are the eigenstates of the operator $A+i\lambda B$ \cite{Barone_2021}, it is straightforward to see that for the above pair of incompatible observables, the operator $\sigma_x+i\lambda \sigma_z$ is nothing but $H(\lambda)$ and one can use $\lambda$ to infer about the states lying on the MUS lines.~However, at the EP $(\lambda=1)$, it is not trivial and requires one to separately check if the state is an eigenstate of $\mathcal{PT}$ operator or not in order to know if the state is $\mathcal{PT}$ symmetric state or $\mathcal{PT}$ broken state.~The MUS within the $G$-metric inner product framework remain the eigenstates of $A+i\lambda_GB$ operator, where $\lambda^2_G=\Delta A^2_G/\Delta B^2_G$, and $A$, $B$ are good observables (see Appendix \ref{appendix C}).~We use these facts to obtain that in the $\mathcal{PT}$-symmetric phase, the MUS line $p=1$ contains the $\mathcal{PT}$ broken states and the lines $\theta=\pi/2$, $3\pi/2$ contain all the $\mathcal{PT}$ symmetric states.~In the $\mathcal{PT}$-broken phase, all the MUS on the lines $p=0$, $p=\infty$ and $\theta=\pi/2$, $3\pi/2$ are $\mathcal{PT}$ broken states and the ones on the lines $\theta=0$, $\pi$, $2\pi$ are $\mathcal{PT}$ symmetric states.~The blue solid lines and red dashed lines in the Fig.~\ref{Figure3} are the MUS lines.

%~It will be more evident in the next section that talks about the arbitrary finite-size system described by a $\mathcal{PT}$-invariant Hamiltonian.

\section{$N\times N$ non-Hermitian Systems}
Indeed, the recipe to derive the condition for good observables in the previous section can be extended to a general $N\times N$ $\mathcal{PT}$-invariant Hamiltonian.
%~The challenge is to find incompatible good observables to construct the uncertainty relation under the $G$-inner product formalism. Interestingly, one good observable for this higher dimensional system is $G$ itself.
%it was explicitly shown that for $2 \times 2$ system, the non-Hermitian Hamiltonian itself is a good observable in the $\mathcal{PT}$-symmetric phase but not in the broken phase.
Also, it is straightforward to show that the non-Hermitian Hamiltonian is a good observable in the $\mathcal{PT}$-symmetric phase but not in the broken phase and  generalize this $2\times2$ result for an arbitrary finite-dimensional $N\times N$ system (see Appendix \ref{appendix A}).~We propose the non-Hermitian Hamiltonian $H$ with the non-zero matrix entries $H_{j,j+1}=H_{j+1,j}=1, ~H_{j,j}=i(-1)^{j+1}\gamma$. 
~Here, $j=1,2,3...N$ for even integers $N$.~This Hamiltonian has been examined by two of the authors previously in~\cite{Modak_2021} and the EP shifts towards $\gamma=0$ as $N$ increases and towards $\gamma=1$ as $N$ approaches towards $2$.

We now construct a Hermitian good observable with the non-zero matrix entries $O_{j,N+1-j}= i(-1)^j$ for even integers $N$.~This good observable $O$ reduces to $\sigma_y$ for $N=2$ and also remains to be a good observable in the $\mathcal{PT}$-broken phase as well.

To characterize the good observables, we now define an operator, $K=O^{\dag}G -GO$ followed by the identifier 
\begin{equation*}
    \kappa=N^{2}\frac{\sum_{rs}|K_{rs}|}{(\sum_{\alpha \beta}|O_{\alpha \beta}|)(\sum_{lm}|G_{lm}|)}.
\nonumber 
\end{equation*}
~The identifier $\kappa$ is defined such that $\kappa=0$ for good observable and non-zero otherwise.~Without any loss of generality, $\kappa$ is defined such that its order of magnitude remains the same for different values of $N$ in the $\mathcal{PT}$-broken phase.~The upper panel of Fig.~\ref{Figure7} shows $\gamma$ vs.~$\kappa$ for different values of $N$.~It is evident that $\kappa=0$ in the $\mathcal{PT}$-symmetric phase characterized by $\gamma <\gamma_{EP}$ for the Hamiltonian $H$.~Thus, $\kappa$ can be used as a parameter to detect the transition from the $\mathcal{PT}$-symmetric phase to the $\mathcal{PT}$-broken phase.
%~Although, for $2\times 2$ systems in the $\mathcal{PT}$-broken phase $H (1/\gamma)$ is a good observable but this is not extendable for the $N\times N$ systems.
~Additionally, Fig.~\ref{Figure7} (inset) also depicts that the Hermitian observables $O$ incompatible with $H$ is also a good observable.
%~Near $\epsilon \to 0$, $\epsilon ^{1/2} $ is larger compared to $\epsilon$ \cite {Yu_2020}.

%\begin{figure}
%\centering
%\includegraphics[width=0.8\linewidth]{unfig4.pdf}
%\caption{Variation of $\kappa$ with $\gamma$ for $H^A$ and $H^B$.~}
%\label{Figure4}
%\end{figure}

%%%%%%%%%%%%%%%%%%%% NEW RESULTS %%%%%%%%%%%%%%%%%%%%%%%%%%%%%%%%%%%%%%%%%%
%\begin{figure}
%\centering
%\includegraphics[width=0.8\linewidth]{unfig5.pdf}
%\caption{Variation of $\kappa$ with $\gamma$ for $H^A$ in case of $O_{1N}$ (main) and $O_{2N}$ (inset).}
%\label{Figure5}
%\end{figure}
%%%%%%%%%%%%%%%%%%%%%%%%%%%%%%%%%%%%%%%%%%%%%%%%%%%%%%%%%%%%%%%%%%%%%%%%%%%%%
\section{Fisher Information gain for MUS:}
%\nam {In this section, we define cumulative gain in the Fisher information with respect to the Hermitian analogue of the system for the incompatible good observables $O$ and Hamiltonian $H$.~We compare the gain in this cumulative Fisher information for MUS with random states.}
Generally, non-Hermitian systems provide for better quantum sensors near the EP.~The degeneracy lift scales with the perturbation ($\epsilon$) for Hermitian vs. non-Hermitian systems as $\epsilon$ vs. $\epsilon ^{1/2}$ \cite {Yu_2020}.~Therefore, it is relevant to compare the Fisher information and thus precision in the actual measurement of the two incompatible observables for the non-Hermitian systems \cite{Huang_2019} with its Hermitian counterpart.

\begin{figure}
\centering
\includegraphics[width=0.95\linewidth]{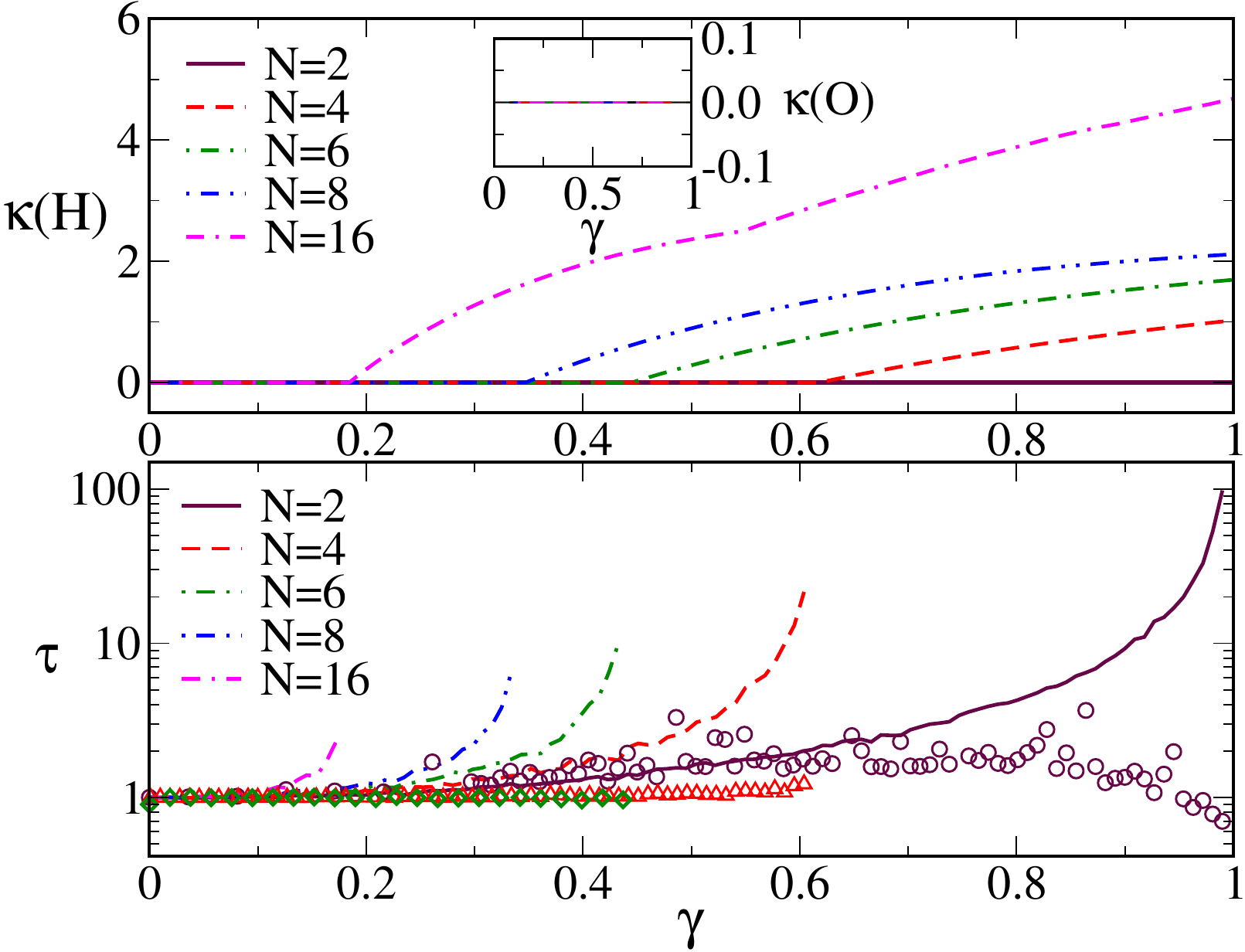}
\caption{The lower panel shows variation of cumulative Fisher information $\tau$ with respect to $\gamma$.~All lines correspond to MUS and symbols correspond to random states.
The upper panel shows $\kappa$ for $H$ (main) and the Hermitian good observable $O$ (inset) with $\gamma$, showing these are good observables.}
\label{Figure7}
\end{figure}

In usual Hermitian Quantum mechanics, QFI for any pure states $\rho=\ket{\psi}\bra{\psi}$ is proportional to the variance of the operator $A$, i.e., $F_Q=4\Delta A^2$.~We define the cumulative gain in the Fisher information in the measurement of incompatible observables $H$ and $O$ by
\begin{equation}
\tau (H,O)=\frac{\Delta H^{2}_G \Delta O^{2}_G}{\Delta \tilde{H}^2 \Delta O^{2}},
\end{equation}
where $\tilde{H}=G^{1/2}HG^{-1/2}$ is the Hermitian counterpart of the non-Hermitian Hamiltonian and the same applies to $O$ via an inverse transform.~It is important to note that the Hermitian counterpart of any good observable is not unique, e.g., if $G=V^{\dagger}V$, then 
$\tilde{O}=VOV^{-1}$ also qualifies as a Hermitian counterpart of $O$.~However, we choose $V=G^{1/2}$ to ensure that the Hermitian counterpart of Hermitian good observable is the observable itself, i.e., for $O=O^{\dagger}$, its Hermitian counterpart $\tilde{O}=O$ (see Appendix \ref{appendix E}).~This quantity $\tau$ quantifies the advantage in the measurement precision of two incompatible observables for Non-Hermitian systems over its Hermitian counterpart.~We plot this gain quantity $\tau$ with the parameter $\gamma$ in the lower panel of Fig.~\ref{Figure7} for MUS and random states, for different $N$ values.~It is evident that the MUS states when averaged over the $\mathcal{PT}$ symmetric states and $\mathcal{PT}$ broken states show remarkable gain near the EP compared to the random states.~Note that the results presented here are averaged over $20000$ random realizations, we have also checked the convergence of the results by varying numbers of random realizations.~One can observe this explicitly for $H$ in Eq.~\eqref{Hamiltonian} and $\sigma_y$ as for the smallest $2\times2$ system in the $\mathcal{PT}$-symmetric  phase.~This implies that the best precision can be achieved near the EP when measuring the observables $O$ with $H$ in the MUS states.~Such a measurement with the proposed candidates of observables in the MUS may lead to the simultaneous non-Hermitian quantum sensors near the EP.
\section{Conclusion}
The Heisenberg uncertainty relation \eqref{Heis-Rob} does not apply to the non-Hermitian systems.~In this letter, we have shown that for any finite-dimensional $\mathcal{PT}$-invariant non-Hermitian systems the observables need not necessarily be Hermitian operators but good observables \eqref{good obs}.~We have constructed the uncertainty relation \eqref{Heis-Rob_G} for these good observables and identified MUS which mark a boundary for the violation of the uncertainty relation.

The fact that the non-Hermitian Hamiltonian itself qualifies a good observable in the $\mathcal{PT}$-symmetric phase can be used as a diagnostic tool to detect the EP for a general $N\times N$ Hamiltonians.~This can be experimentally realized in an ultracold fermionic system as a Hamiltonian with gain and loss term.

The proposed non-Hermitian uncertainty relation has a potential application in the field of quantum metrology.~We found that the cumulative Fisher information gain increases near the EP when compared with the corresponding Hermitian systems.Also, the MUS are shown to be the best candidate states for measuring two incompatible good observables with enhanced precision, one of them being the non-Hermitian Hamiltonian.~We proposed to apply this feature to realize non-Hermitian quantum sensing near the EP, leading to the intelligent quantum sensors.
%It is of great scope to look at the time evolution dynamics of non-Hermitian systems in the context of uncertainty relation and also the behaviour of tighter sun uncertainty relation for such systems.
\section{Acknowledgements}
RM acknowledges the DST-Inspire fellowship by the Department of Science and Technology, Government of India, SERB start-up grant (SRG/2021/002152), and IIT Tirupati CPDA grant for their support.~NS acknowledges the IOE seed grant by the Banaras Hindu University (Seed Grant II/2021-22/39995). BPM acknowledges the research grant for faculty under IoE Scheme (Number 6031) of Banaras Hindu University, Varanasi.~The authors acknowledge Arun K Pati,  Aravinda S., and Francisco M. Fernandez for fruitful discussions.  
\appendix

\section{Is Hamiltonian a good observable?}\label{appendix A}
In this section, we prove that for any arbitrary finite dimensional system, the non-Hermitian Hamiltonian itself is a good observable in the $\mathcal{PT}$-symmetric 
phase but not in the broken phase.~In order to find whether the non-Hermitian Hamiltonian $H$ 
is a good observable or not, one needs to check whether $H$ satisfies the condition of good observable, i.e., $H^{\dag}G=GH$ , where $G=\sum_i\ket{L_i}\bra{L_i}$. $\ket{L_i}$ and $\ket{R_i}$ are left eigenvector and right eigenvector of $H$, and they satisfy the eigenvalue equations $H^{\dag}\ket{L_i}=h_i\ket{L_i}$ and $H\ket{R_i}=h^*_i\ket{R_i}$, respectively.
The vectors $\ket{L_i}$ and $\ket{R_i}$ also satisfy bi-orthonormalization condition, i.e., $\la R_i|L_j\ra=\delta_{ij}$ and it automatically implies
$\sum_i\ket{R_i}\bra{L_i}=\mathds{1}$.~All $h_i$s are completely real ($h_i=h^*_i$) when $H^{\dag}$ (or $H$) belongs to the $\mathcal{PT}$-symmetric phase
and in the broken phase at least two of the eigenvalues must be complex. Now,
\begin{equation}
    H^{\dag}G=\sum_iH^{\dag}\ket{L_i}\bra{L_i}=\sum_i h_i \ket{L_i}\bra{L_i}
    \label{goH1}
\end{equation}
On the other hand, 
\begin{align}
    GH&=\sum_i\ket{L_i}\bra{L_i}H=\sum_i \ket{L_i}\bra{L_i}H\sum_j\ket{R_j}\bra{L_j} \nonumber \\
    &=\sum_{ij}h^{*}_j
    \ket{L_i}\bra{L_j}\delta_{ij}=\sum_i h^{*}_i \ket{L_i}\bra{L_i}
    \label{goH2}
\end{align}
It says that $H^{\dag}G=GH$ will be satisfied only if $h_i=h^*_i$, i.e., $H$ belongs to the $\mathcal{PT}$-symmetric phase.~On the 
other hand in the broken phase, the following condition can not hold. Hence, $H$ will be always a good observable in the symmetric phase but not in the broken phase.

\section{Uncertainty relation for good observables} \label{appendix B}
We prove the uncertainty relation for the $G$-inner product and good observables in this section. In the usual quantum mechanics where the observables are Hermitian, the Heisenberg Uncertainty relation for two incompatible observables $A$ and $B$ is given by 
\begin{equation}
\Delta A^2\Delta B^2\geq \frac{1}{4}{\vert\bra{\psi}[A,B]\ket{\psi}\vert}^2,  
\label{Heis-Rob-A}
\end{equation}
where $|\psi\ra$ is a state in the Hilbert space, and $\Delta A^2=\la \psi|A^2|\psi\ra -(\la \psi|A|\psi\ra)^2$, $\Delta B^2=\la \psi|B^2|\psi\ra -(\la \psi|B|\psi\ra)^2$. 
On relaxing the \quotes{Hermiticity} condition the variances are given by
\begin{align}
\Delta A^2&=\la \psi|A^{\dag}A|\psi\ra -\la \psi|A^{\dag}|\psi\ra \la \psi|A|\psi\ra, \nonumber\\ 
\Delta B^2&=\la \psi|B^{\dag}B|\psi\ra -\la \psi|B^{\dag}|\psi\ra \la \psi|B|\psi\ra, \nonumber\\
\label{var-nonH}
\end{align}
and the commutator term on the RHS of Eq.~\eqref{Heis-Rob-A} can be written as
\begin{align}
&\frac{1}{4}|\la \psi |[A,B]|\psi\ra|^2\nonumber\\
&=\frac{1}{4}\la \psi |(AB-BA)|\psi\ra \la \psi |(B^{\dag}A^{\dag}-A^{\dag}B^{\dag})|\psi\ra. 
\end{align}
In the $G$-inner product formalism the adjoint the operators and the inner product need to be redefined \cite{Ju_2019} under the rule $A^{\dag}\to G^{-1}A^{\dag}G$, $B^{\dag}\to G^{-1}B^{\dag}G$ and the inner product under $\la \psi |Q|\psi \ra \to \la \psi |GQ|\psi \ra$. 
LHS and RHS of the Eq.~\ref{var-nonH} can now be written as
\begin{align}
\Delta A_G^2&=\la \psi|A^{\dag}GA|\psi\ra -\la \psi|A^{\dag}G|\psi\ra \la \psi|GA|\psi\ra, \nonumber\\
\Delta B_G^2&=\la \psi|B^{\dag}GB|\psi\ra -\la \psi|B^{\dag}G|\psi\ra \la \psi|GB|\psi\ra, \nonumber\\
\end{align}
\begin{align}
&\frac{1}{4}|\la \psi |G[A,B]|\psi\ra|^2 \nonumber\\
& =\frac{1}{4}\la \psi |G(AB-BA)|\psi\ra \la \psi |(B^{\dag}A^{\dag}-A^{\dag}B^{\dag})G|\psi\ra.
\end{align}
Additional constraint of $A$ and $B$ 
being good observables, i.e., $A^{\dag}G=GA$ and $B^{\dag}G=GB$ leads to
\begin{align}
\Delta A_G^2&=\la \psi|GA^2|\psi\ra -(\la \psi|GA|\psi\ra)^2, \nonumber\\
\Delta B_G^2&=\la \psi|GB^2|\psi\ra -(\la \psi|GB|\psi\ra)^2, \nonumber\\
\nonumber\\
\end{align}
\begin{align}
&\frac{1}{4}|\la \psi |G[A,B]|\psi\ra|^2\nonumber\\
&=-\frac{1}{4}(\la \psi |G(AB-BA)|\psi\ra)^2= [\frac{1}{2i}\la \psi |G[A,B]|\psi\ra]^2.
\end{align}
For the uncertainty relation to hold for the two good observables $A$ and $B$, the following inequality must satisfy
\begin{equation}
\Delta A_G^2 \Delta B_G^2\geq \frac{1}{4}\vert\bra{\psi}G[A,B]\ket{\psi}\vert^2,  
\label{G-inner A}
\end{equation}
where $\Delta A_G^2=\la \psi|GA^2|\psi\ra -(\la \psi|GA|\psi\ra)^2$ and
$\Delta B_G^2=\la \psi|GB^2|\psi\ra -(\la \psi|GB|\psi\ra)^2$.\\

It is straightforward to show that the above inequality holds true for good observables in the following subsequent steps. Let us define two vectors in a vector space 
\begin{align}
\ket{f}&=(A-\la \psi|GA|\psi\ra)|\psi\ra, \nonumber\\
\ket{g}&=(B-\la \psi|GB|\psi\ra)|\psi\ra.
\end{align}
The dual vectors are
\begin{align}
\bra{f}=\la \psi| (A^{\dag}-\la \psi|A^{\dag}G|\psi\ra),\nonumber\\
\bra{g}=\la \psi | (B^{\dag}-\la \psi|B^{\dag}G|\psi\ra).
\end{align}
Given that $A$ and $B$ are good observables, $\la \psi|GA^{\dag}|\psi\ra, \la \psi|GA|\psi \ra, \la \psi|GB^{\dag}|\psi\ra, \la \psi|GB|\psi \ra$ are all real quantities.~On plugging in $|f\ra$, and $|g\ra$ in the Cauchy–Schwarz inequality,
 \begin{align}
 \la f|G|f\ra\la g|G|g\ra&\geq \left\vert\la f|G|g\ra\right\vert^2 \nonumber\\ &\geq\frac{1}{2}\left\vert\la f|G|g\ra - \la g|G|f\ra\right\vert^2. 
 \label{C-S_fg_inq}
 \end{align}
Different terms in the above inequality can be simplified to 
\begin{align}
\la f|G|f\ra&=\la \psi|(A^{\dag}-r')G(A-r)|\psi\ra  =  \la \psi|GA^2|\psi\ra  -r^2, \nonumber \\
\la g|G|g\ra&=\la \psi|(B^{\dag}-p')G(B-p)|\psi\ra  =  \la \psi|GB^2|\psi\ra - p^2,\nonumber \\
\la f|G|g\ra&= \la \psi|(A^{\dag}-r')G(B-p)|\psi\ra=\la \psi|GAB|\psi\ra -rp, \nonumber \\
     \la g|G|f\ra&= \la \psi|(B^{\dag}-p')G(A-r)|\psi\ra  =\la \psi|GBA|\psi\ra -rp, \nonumber
\end{align}
for $r=\la \psi |GA|\psi \ra$, $r'=\la \psi |GA^{\dag}|\psi \ra$, $p=\la \psi |GB|\psi \ra$, $p'=\la \psi |GB^{\dag}|\psi \ra$ and $\la \psi |G|\psi \ra =1$.~For the real $r,p$ and $(GAB)^{\dag}=GBA$ we have $\la f|G|g\ra ^{*}=\la g|G|f\ra$. Collecting these relation in Eq.~\ref{C-S_fg_inq} leads to Eq.~\ref{G-inner A}.\\

Therefore, the Heisenberg's uncertainty relation holds true for $G$-metric inner product and good observables. However, the following derivation does not go through for any arbitrary operators and the violation of the inequality in Eq.~\ref{G-inner A} can be seen for not good observables as shown in the main text. 

 \section{Minimum Uncertainty states (MUS)}\label{appendix C}
 On the lines of a result in the usual quantum mechanics\cite{Barone_2021}, one would expect that for the $G$-inner product formalism and good observables $A$ and $B$, the MUS are the eigenstates of operator $A+i\lambda_G B$. We prove it here in this section.~The minimum uncertainty states in the $G$-metric formalism and good observables saturate the inequality in Eq.~\ref{G-inner A}, i.e.,\\
 \begin{equation}
 \Delta A_G^2 \Delta B_G^2= \frac{1}{4}\left\vert\la \psi |G[A,B]|\psi\ra\right\vert^2 
 \label{MUS-G_metric}
 \end{equation}
 In the context of Cauchy-Schwarz inequality for any two vectors the above equality corresponds to 
\begin{align}
\la f|G|f\ra\la g|G|g\ra=|\la f|G|g\ra|^2= \big(\frac{1}{2}\left\vert\la f|G|g\ra-\la g|G|f\ra\right\vert\big)^2. 
\end{align}
This implies
\begin{align}
&\frac{1}{2}\left\vert\la f|G|g\ra + \la g|G|f\ra\right\vert^2 =0, \nonumber \\
&\la \psi|GAB|\psi \ra + \la \psi|GBA|\psi \ra =2rp. \nonumber \\
\label{mus1}
\end{align}
 Separately, one can infer that $|f\ra =\mu |g\ra$ from $\la f|G|f\ra\la g|G|g\ra=|\la f|G|g\ra|^2$ for a complex number $\mu$.~Therefore, 
 \begin{align}
 (A-r)|\psi\ra=\mu (B-p)|\psi\ra. \label{mus2}
 \end{align}
Using the conditions in Eq.~\ref{mus1} and Eq.~\ref{mus2}, one can show that MUS are the eigenstates of the operator $A+i\lambda_G B$, i.e, 
\begin{equation}
(A+i\lambda_G B)|\psi\ra=(r+i\lambda_G p)|\psi\ra,    
\label{mus3}
\end{equation}
 where $\lambda_G$ is a real number given by 
 \begin{equation*}
 \lambda_G^{2}=(\la \psi|GA^2|\psi\ra -r^2)/(\la \psi|GB^2|\psi\ra -p^2).    
 \end{equation*}
 Note that $\la \psi|GB^2|\psi\ra \neq p^2$.\\
 
Hence, for the $G$-inner product formalism and good observables the MUS condition is perfectly aligned with the usual quantum mechanics for $G=\mathds{1}$.\\
 
It is possible to infer some MUS discussed for specific cases in the main text, from the above result.~We have shown in Fig.4 (in the main text) that the general state $\mathcal{N}\left(|E_1\ra+pe^{i\theta}|E_2\ra\right)$ are MUS for $p=1$ and $\theta=\pi/2$, $3\pi/2$, for the operators 
$A=H(\gamma)$, and $B=\sigma_y$ in the $\mathcal{PT}$-symmetric phase. Here $\mathcal{N}$ is the normalization constant such that $\la \psi|G|\psi\ra=1$. Note that both $A$ and $B$ here are good observables as per our definition.~At least for $\theta=\pi/2$, i.e., $|\psi\ra=\mathcal{N}(|E_1\ra+i|E_2\ra)$, it is straightforward to check that indeed the state is one of the MUS in the following steps.
\begin{align}
&[H(\gamma)+i\lambda_G \sigma_y]\mathcal{N}\left(|E_1\ra+i|E_2\ra\right)\nonumber\\
&=\mathcal{N}\left(\epsilon |E_1\ra-i\epsilon |E_2\ra\right)+
i\lambda_G \sigma_y\mathcal{N}\left(|E_1\ra+i|E_2\ra\right), 
\end{align}
where $\pm\epsilon$ are the eigenvalues of $H(\gamma)$. 
Since $\sigma_y|E_1\ra=-|E_2\ra$ and $\sigma_y|E_1\ra=-|E_2\ra$, $\mathcal{N}(|E_1\ra+i|E_2\ra)$ is an eigenstate of $[H(\gamma)+i\lambda_G \sigma_y]$ if $\epsilon=-\lambda_G$ and 
\begin{equation*}
\lambda_G^{2}=\epsilon^2=\frac{(\la \psi|GA^2|\psi\ra -r^2)}{(\la \psi|GB^2|\psi\ra -p^2)}.
\end{equation*}\\
Therefore, the condition for the state $|\psi\ra=\mathcal{N}(|E_1\ra+i|E_2\ra)$ to be one of the MUS is $\epsilon=-\lambda_G$. In fact, the condition for the state $\mathcal{N}(|E_1\ra+pe^{i\theta}|E_2\ra)$ to be MUS is $\epsilon= -\lambda_G \sin\theta$. 
%It is important to mention that these results nicely fall in place for the Dirac inner product for $G=\mathds{1}$.  

%%%%%%%%%%%%%%%%%%%%%%%%%%%%%%%%%%%%%%%%%%%%%%%%%%%%%%%%%%%%%%%%%%%%%%%%%%%%%%%%%%%%
  \section{MUS: symmetry of the state} \label{appendix D}
  We have observed for the Dirac 
  inner product formalism and the observables $A=\sigma_x$ and $B=\sigma_z$ that the generic state $\mathcal{N}\left(|E_1\ra+pe^{i\theta}|E_2\ra\right)$ corresponds to the lines $p=1$, $0$, $\infty$ (for all values of $\theta$) or $\theta=0$, $\pi$, $2\pi$ (for all values of $p$) in the $\mathcal{PT}$-symmetric phase. 
 
 It is established that these MUS must be eigenstates of the operator $A+i\lambda B$,  where $\lambda=\Delta A/\Delta B$.~If one chooses $A=\sigma_x$ (or $\sigma_y$), and $B=\sigma_z$, then $A+i\lambda B$ will itself be non-Hermitian $\mathcal{PT}$-invariant operator with all the real eigenvalues for $\lambda <1$ and complex eigenvalues for $\lambda >1$.~Therefore, it is self explanatory that for given MUS if $\lambda<1(>1)$ the MUS are $\mathcal{PT}$ symmetric (broken) states.~However, for $\lambda=1$ nothing can be concluded from this analysis. One such example is $\mathcal{N}\left(|E_1\ra+|E_2\ra\right)$ in the $\mathcal{PT}$-symmetric phase.~One needs to check separately if such states are eigenstates of the $\mathcal{PT}$ operator. 

\section{Hermitian counterpart of good observable for $2\times 2$ system } \label{appendix E}
As we discussed in the main text, the Hermitian counterparts of good observables in a particular $G$-metric space is not unique.~Given $G=V^{\dagger}V$, 
$\tilde{O}=VOV^{-1}$ is Hermitian counterpart of the good observable $O$.~In this work, we choose $V=G^{1/2}$. For the $2\times 2$ Hamiltonian, it is straightforward to check that in the $\mathcal{PT}$-symmetric phase,  
\begin{equation}
G^{1/2}=
\begin{pmatrix}
(p_1+p_2)/2 & i(p_1-p_2)/2 \\
-i(p_1-p_2)/2 & (p_1+p_2)/2  \\
\end{pmatrix},
\label{gs_appen}
\end{equation}
where $p_1=(\frac{1+\gamma}{1-\gamma})^{1/4}$ and $p_2=(\frac{1-\gamma}{1+\gamma})^{1/4}$. Now, the Hermitian counterpart of the good observable $\sigma_y$ (Hermitian) in the $G$-metric space can be obtained as  $\tilde{\sigma_y}=G^{1/2}\sigma_yG^{-1/2}=\sigma_y$.~
On the other hand, if we choose 
\begin{equation}
V=
\begin{pmatrix}
\frac{1}{(1-\gamma^2)^{1/4}} & \frac{-i\gamma}{(1-\gamma^2)^{1/4}} \\
0 & (1-\gamma^2)^{1/4}  \\
\end{pmatrix},
\end{equation}
such that $G=V^{\dagger}V$, it turns out that $\tilde{\sigma_y}=V\sigma_yV^{-1}=\gamma \sigma_z +(1-\gamma^2)^{1/2}\sigma_y \neq \sigma_y$. 
It shows that for our choice of $V=G^{1/2}$, the Hermitian counterpart of the Hermitian good observable remains the same, which is not the case in general for other $V$.~This remains true for the $N\times N$ Hamiltonian as well.~Our choice of transformation is better to study the comparative gain in the Fisher information.

\section{Time-dependent good observable in the $\mathcal{PT}-$ broken phase} \label{appendix F}
As we discussed in the main text, the $G$-metric is time-dependent in the $\mathcal{PT}$-broken phase.~This also makes the good observables in general time-dependent.~In the main text, we have constructed the good observable at the initial time $t=0$.~However, the time-dependent $G$-metric in the broken phase is given by~\cite{Huang_2021}, 
\begin{equation}
G^{b}(t)=\frac{1}{\Lambda}
\begin{pmatrix}
g_{11}& g_{12}\\
 g_{21} & g_{22} \\
\end{pmatrix},
\label{gb_appen}
\end{equation} where $\Lambda=\sqrt{\gamma^2-1}$, $g_{11}=\gamma \cosh(2t\Lambda)-
\Lambda\sinh(2t\Lambda)$, $g_{22}=\gamma \cosh(2t\Lambda)+\Lambda\sinh(2t\Lambda)$, $g_{12}=g^*_{21}=-i\cosh(2t\Lambda)$. Now, it is straightforward to check using the identity 
$O^{\dagger}(t)G^{b}(t)=G^{b}(t) O(t)$ that the Hermitian operator 
\begin{equation}
O_1(t)=
\begin{pmatrix}
-\Lambda\tanh(2t\Lambda)& -i\\
 i & \Lambda\tanh(2t\Lambda) \\
\end{pmatrix},
\label{gb_appen_1}
\end{equation}
is a good observable.~Note that $O_1(t=0)=\sigma_y$.~Additionally, one can also see that the time-dependent non-Hermitian operator
\begin{equation}
O_2(t)=
\begin{pmatrix}
i\frac{s_1}{\gamma s_1 -\Lambda s_2} & \frac{\gamma s_1 +\Lambda s_2}{\gamma s_1 -\Lambda s_2}  \\
1 & -i\frac{s_1}{\gamma s_1 -\Lambda s_2} \\
\end{pmatrix},
\label{gb_appen_1}
\end{equation}
where $s_1=\cosh(2t\Lambda)$ and $s_2=\sinh(2t\Lambda)$, also satisfies the good observable condition.~In this case,  $O_2(t=0)=H(1/\gamma)$, which is indeed our $2\times2$ good observables in the $\mathcal{PT}$-broken phase.

\bibliography{ptuncertainty}
 \end{document}